\documentclass[aps,prl,reprint,groupedaddress]{revtex4-1}
\usepackage{graphicx}
\usepackage{dcolumn}
\usepackage{bm}
\usepackage{amssymb}
\usepackage{amsmath}
\usepackage{amsfonts}
\usepackage{epsf}
\usepackage{epsfig}
\usepackage{color}
\usepackage{siunitx}
\usepackage{epstopdf}

\begin{document}
\title{
\large
Spin and angular momentum in strong-field ionization}

\author{D.~Trabert} 
\email{trabert@atom.uni-frankfurt.de}
\author{A.~Hartung} 
\author{S.~Eckart} 
\author{F.~Trinter}
\author{A.~Kalinin}
\author{M.~Sch\"offler}
\author{L.~Ph.~H.~Schmidt}
\author{T.~Jahnke}
\author{M.~Kunitski}
\author{R.~D\"orner}
\email{doerner@atom.uni-frankfurt.de}

\affiliation{
 Institut f\"ur Kernphysik, Goethe-Universit\"at, Max-von-Laue-Str.~1, 60438 Frankfurt, Germany \\
}

\date{\today}

\begin{abstract}
The spin polarization of electrons from multiphoton ionization of Xe by 395~nm circularly polarized laser pulses at $6\cdot10^{13}$~W/cm$^2$ has been measured. At this photon energy of 3.14~eV the above threshold ionization peaks connected to Xe$^+$ ions in the ground state ($J=3/2$, ionization potential $I_p=12.1$~eV) and the first exicted state ($J=1/2$, $I_p=13.4$~eV) are clearly separated in the electron energy distribution. These two combs of ATI peaks show opposite spin polarizations. The magnitude of the spin polarization is a factor of two higher for the $J=1/2$ than for the $J=3/2$ final ionic state. In turn the data show that the ionization probability is strongly dependent on the sign of the magnetic quantum number.   
\end{abstract}

\maketitle

Light-driven ionization processes are sensitive to the spin of the electron. Surprisingly, this important and fundamental fact of light matter interaction is experimentally well validated only for the special cases of single photon and resonant enhanced two and three photon processes \cite{Granneman76}. For strong-field ionization it rests on only one single experiment \cite{Hartung2016}, which did not even resolve the quantum state from which the electron was ejected. 

The role of the spin in single photon ionization was adressed soon after the discovery of the electron spin \cite{Sauter31}. Starting in the 1960s, it became clear that spin selectivity of single photon ionization of atoms and molecules is very general. Today it is well studied experimentally and theoretically (see \cite{Cherepkov83} for a review). The generalization to the multiphoton regime was achieved first in pioneering theoretical work by Lambropoulus \cite{Lambropoulos73}. Recently Barth and Smirnova \cite{Barth2013} predicted a high degree of spin polarization for strong-field ionization by circularly polarized femtosecond pulses. The proposed mechanism giving rise to spin sensitivity of strong-field ionization consists of two independent steps. The primary effect is that the nonadiabatic tunneling probability through a rotating barrier depends on the sign of the magnetic quantum number $m_{l}$ of the orbital. This finding was confirmed experimentally \cite{Herath12} and by solving the time-dependent Schr\"odinger equation \cite{BarthLein2014} without invoking the concept of tunneling explicitly. Together with the strong binding energy dependence of strong-field ionization the $m_{l}$ dependence then leads to a spin selectivity. Because, due to the spin-orbit interaction, the binding energy differs for parallel or antiparallel orientation of the spin with respect to the projection of the orbital angular momentum $m_l$ on the quantization axis.

\begin{figure}[t]
\centering
\epsfig{file=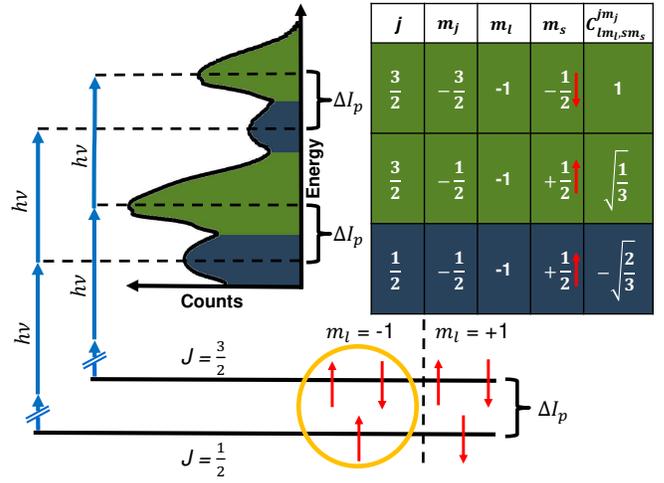, width=1.\columnwidth} 

\caption{Schematics of multiphoton ionization in xenon by $395$~nm ($h\nu=3.14$~eV) laser pulses. The ground state ($^2P$, $J=3/2$) and first excited state ($^2P$, $J=1/2$) of Xe$^+$ differ by $\Delta I_{p}=1.3$~eV in ionization potential. Multiphoton ionization therefore yields two combs of peaks spaced by the photon energy with a relative offset of $1.3$~eV. If only electrons from the orbital with $m_{l}=-1$ (yellow circle) are ejected one obtains a spin polarization $S_{1/2}=+1$ for the $J=1/2$ final state and $S_{3/2}=-0.5$ for the $J=3/2$ final state (the ground state). $C_{lm_{l},sm_{s} }^{jm_{j}}$ denoting the relevant Clebsch-Gordan coefficients leading to $S_{1/2}=-2\cdot S_{3/2}$.} 
\label{fig1}
\end{figure}

The purpose of the present paper is to experimentally show this theoretically suggested connection of spin, magnetic quantum number and binding energy in strong-field ionization. This relies on experimentally determining both: the ionization potential and the spin polarization of the electron. For xenon this is possible as illustrated in Fig. \ref{fig1}. Strong-field ionization by a laser pulse with a photon energy $h\nu=3.14$~eV i.e. a wavelength of $\lambda=395$~nm leads to a comb of peaks in the electron energy ($E_e$) distribution, equally spaced by the photon energy: 
\begin{eqnarray}
E_e = nh\nu - I_p- 2U_p \label{eqati}
\label{eq1}
\end{eqnarray}  
Here $n$ is the number of absorbed photons, $I_p$ the ionization potential for the respective ionic final state and $U_p$ is the ponderomotive potential at the given laser intensity $I$ ($U_p=0.44 \pm 0.14$~eV, Keldysh parameter $\gamma=3.73 \pm 0.80$). These maxima are referred to as above threshold ionization (ATI) peaks. For xenon removing an outer electron with its spin parallel to its orbital angular momentum ($j=3/2$) yields the ionic ground state $^2P$ ($I_p=12.1$~eV with total angular momentum $J=3/2$) while emission of an electron with opposite spin ($j=1/2$) leads to the first excited state of the ion ($J=1/2$). The ionization potential for this case is higher by $1.306$~eV. Thus at a photon energy of $3.14$~eV the two combs of ATI peaks belonging to the two different ionic states ($J=1/2$ and $J=3/2$) do not overlap in energy. The mechanism responsible for the spin selectivity predicts a sign change of the spin for electrons from these two different ATI combs. 

As can be seen in Fig. \ref{fig1} there are six relevant electrons in the outermost shell of xenon (the ionization from states with $m_{l}=0$ is strongly suppressed \cite{BarthSmirnova11} and therefore neglected). For the two electrons with $j=1/2$ total angular momentum, the magnetic quantum number $m_{l}$ and spin orientation $m_{s}$ are directly intertwined, yielding $m_{s}=-1/2$ for $m_{l}=+1$ and $m_{s}=+1/2$ for $m_{l}=-1$. The same mechanism applies to the four electrons with $j=3/2$, although the situation is more complicated. Both spin orientations are possible for $m_{l}=+1$ and $m_{l}=-1$, respectively. The net spin polarization $S_{3/2}$ results from different Clebsch-Gordan coefficients $C_{lm_{l},sm_{s} }^{jm_{j}}$ for $|m_{j}|=1/2$ and $|m_{j}|=3/2$ (inset in Fig. \ref{fig1}). As a result $S_{1/2}$ and $S_{3/2}$ follow the relation $S_{1/2}=-2\cdot S_{3/2}$.     

\begin{figure}[t]
\centering
\epsfig{file=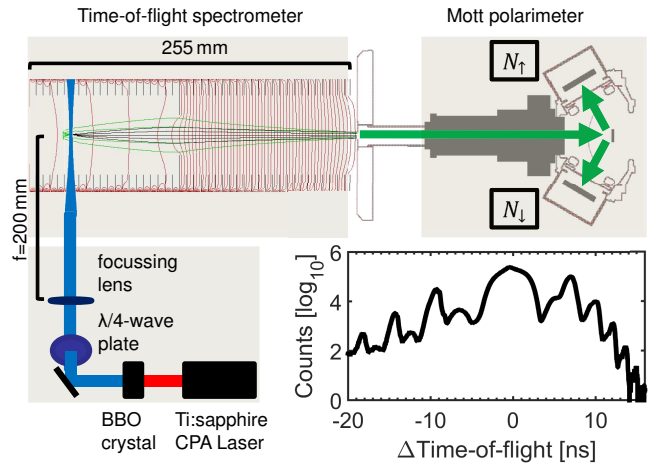, width=1.\columnwidth} 
\caption{TOF-Mott spectrometer. Electrons created in the laser focus are guided in a spectrometer by a combination of a weak acceleration field ($7.5$~V/cm), a focussing lens and high field region ($68$~V/cm) into the entrance hole of a commercial Mott polarimeter. After backscattering at a gold target, the electrons are detected by four micro-channel plate detectors, two placed in the plane of the figure (as shown) and two normal to the plane of the figure (not shown). The inset shows a time-of-flight spectrum relative to zero momentum on one of the detectors (positive values correspond to electrons starting towards the detector). The energy resolution is best for electrons with negative delta time-of-flight. Only this part of the time-of-flight spectrum is used for the analysis.}
\label{fig2}
\end{figure}

We used a ${\beta}$-barium borate (BBO) crystal to frequency double $790$~nm laser pulses from a KMLabs Wyvern 500 Ti:sapphire Chirped Pulse Amplification (CPA) Laser system ($40$~fs FWHM, $100$~kHz). A $\lambda/4$-wave plate was used to switch to circularly polarized light with a wavelength of $395$~nm. The light was focused into a xenon gas target using a lens with $f=200$~mm, resulting in a focal averaged intensity of $(6 \pm 2) \cdot 10^{13}$~W/cm$^2$. The intensity was calibrated using the energy shift of ATI peak positions due to the intensity dependent change of the ponderomotive potential according to equation \ref{eqati}. To cancel out instrumental asymmetries e.g. different detector efficiencies, the measurement was performed inverting the helicity of the light every five minutes. The total data acquisition time was 14 hours.   

The emitted electrons travelled through a time-of-flight spectrometer to a Mott spin polarimeter. In order to enhance the accepted solid angle, two regions of different electric field strengths were applied to create an electrostatic lens as shown in Fig. \ref{fig2}. The Mott polarimeter is based on the design in \cite{Burnett94}. In the polarimeter the electrons are accelerated to about $27$~keV onto a gold target. Those which are backscattered are detected by four micro-channel plate detectors, two placed in the plane of light polarization and two normal to the plane of polarization. The measured time-of-flight is given by the sum of the flight times in the spectrometer and in the polarimeter. The latter is calculated to be $12$~ns, almost independent of the initial momentum upon ionization. The time-of-flight spectra recorded with the two opposite micro-channel plate detectors in polarization plane were then used to calculate the measured spin polarization:

\begin{eqnarray}
S=\frac{1}{S_{eff}}\frac{N_{\uparrow}-N_{\downarrow}}{N_{\uparrow}+N_{\downarrow}}
\label{eq2}
\end{eqnarray}  
with the instrumental scaling factor $S_{eff}$ dependent on the detector geometry, the kinetic energy of the scattering electrons and the target material. The value of $S_{eff}=-0.15$ was taken from \cite{McClelland89}.

\begin{figure}[t]
\centering
\epsfig{file=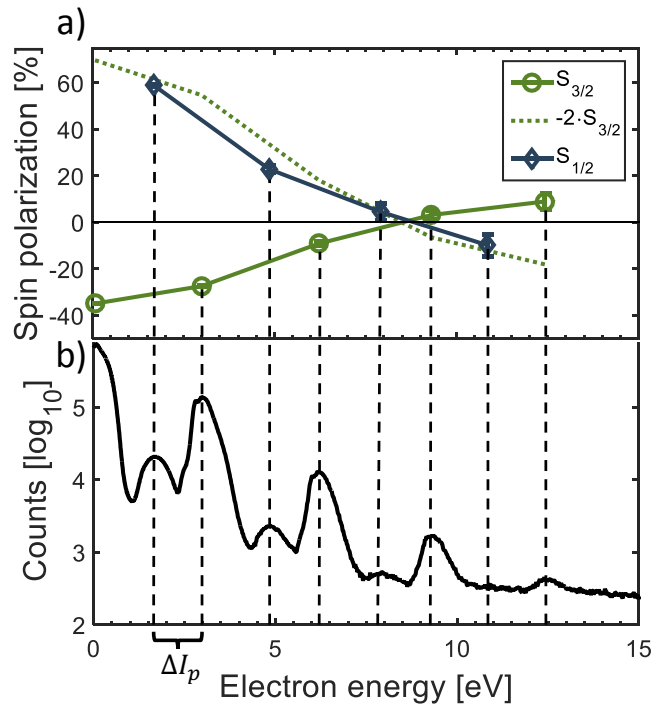, width=1.\columnwidth} 
\caption{Energy-dependent spin polarization $S$ for ionization of xenon by circularly polarized laser pulses ($395$~nm, $40$~fs, $6 \cdot10^{13}$~W/cm$^2$. a) Measured spin polarization for the $J=3/2$ (green circles) and $J=1/2$ (blue symbols) final states of Xe$^+$. The full lines are drawn to guide the eye. The green dotted line is obtained by flipping the green line at the horizontal axis and stretching it by a factor of two vertically. This is what one would expect based on the Clebsch-Gordan coefficients for the two states (see also Fig. \ref{fig1}). b) Electron energy distribution measured on one of the detectors. The comb of high peaks corresponds to the $J=3/2$ final state, the comb of lower peaks is shifted by $1.3$~eV and corresponds to the $J=1/2$ final state of Xe$^+$. The spectrum is not corrected for the energy-dependent solid angle of the time-of-flight spectrometer. Therefore the lowest energies are exaggerated by about a factor of $200$ as compared to the highest energies.}  
\label{fig3}
\end{figure}

Fig. \ref{fig3}(b) shows the measured electron energy distribution. As expected, two combs of ATI peaks, offset by $1.3$~eV are visible. The peak at around $3$~eV electron energy corresponds to an absorption of $5$ photons and belongs to the $J=3/2$ final ionic state. The small peak at around $1.7$~eV corresponds to the same number of absorbed photons but the $J=1/2$ ionic state. The envelope of this histogramm is significantly deformed by a strongly energy-dependent collection solid angle of our spectrometer. From simulation we estimate this solid angle to be around $0.21$~sr at $3$~eV and $0.035$~sr at $10$~eV. From the four detectors in our polarimeter we simultaneously obtain four such electron spectra. Two of those are sensitive to the spin component along the light propagation direction and two are sensitive to the spin polarization in the plane of light polarization. As expected the latter two show no spin polarization and are used for cross checks. 

The resulting spin polarization along the light propagation is depicted in Fig. \ref{fig3}(a). The data show a maximum spin polarization of $60\%$ for the electrons from the  $J=1/2$ state. As expected from the scenario shown in Fig. \ref{fig1} the spin polarization inverts between the $J=1/2$ and $J=3/2$ states. To further test this proposed scenario we show its experimental estimation by the dotted green line. As discussed above, if the spin polarization is caused by the sign of $m_{l}$ dependence of the ionization, one obtains that the spin polarization $S_{1/2}$ is given by $S_{1/2}=-2\cdot S_{3/2}$ (Fig. \ref{fig1}). The excellent agreement of the dotted green line with the blue line based on the measured $S_{1/2}$ thus directly confirms the sign of $m_{l}$ dependence of the ionization process by circularly polarized laser pulses.

It is obvious from the figure that failing to resolve the final ionic state as in \cite{Hartung2016} leads to a much reduced apparent polarization because of two reasons. Firstly, the polarization is reduced by a factor of two originating from the most abundant $J=3/2$ state and secondly the contribution from the $J=1/2$ state showing the opposite sign leads to a partial cancellation of the net polarization.   

Calculation by Barth and Smirnova \cite{Barth2013} predict an increase of the electron energy-integrated spin polarization with rising $\gamma$. Using equations (14)-(16) in \cite{Barth2013} one can expect a relative increase of 43\% between the experiment reported in \cite{Hartung2016} at $\gamma=1.24$ and the present work at $\gamma=3.73$. The strong but not completely known energy dependence of the detection solid angle of our spectrometer does not allow us to obtain a reliable energy-integrated spin polarization, thus a test of this prediction remains a goal for future experimental work.

In conclusion, we have found a strong electron spin polarization for ionization by a strong laser field which has opposite signs for the $J=1/2$ and $J=3/2$ final states of the ion. This observation validates theoretical prediction that the spin polarization in strong-field ionization is a result of the dependence of the ionization on the sign of the magnetic quantum number. One can turn this argument around saying that our experiment provides direct experimental proof of the predicted dependence of strong-field ionization on the sign of the magnetic quantum number. The observed maximum spin polarization of $35\%$ for the $J=3/2$ state corresponds to a $70\%$ difference of the ionization rate between the energetically degenerate $m_{l}=+1$ and $m_{l}=-1$ orbitals. Despite this huge effect, the $m_{l}$ dependence is rarely discussed in strong-field experiments today. The observed pronounced energy dependence shows that not only ionization probability, but also the initial momentum distribution of the electron upon ionization depends on the magnetic quantum number. One application of the observed $J$-state-resolved spin polarization is the creation of a ring current in the ion on an ultrafast time scale as highlighted by \cite{BarthSmirnova14}. Taking our findings further one can also expect related effects in molecular ionization processes. First indications for this have been found in theory \cite{Liu16}. We envision that a particularly exciting application of such spin- and final-state-resolved experiments will be strong-field ionization of chiral molecules. In this case in addition to the sense of rotation of the electronic wave function's phase also the spatial structure of the potential has a handedness. Chirality significantly influences strong-field ionization as has been shown both theoretically \cite{Dreissigacker2014} and experimentally \cite{Beaulieu16}\cite{Beaulieu17}. Furthermore the interplay of spin and chirality has recently found much attention \cite{Michaeli16}. On the more technical level one can envision to use the spin polarized electron flux in a rescattering scenario for attosecond probing the parent ion as suggested in \cite{Milosevic16}. While rescattering is suppressed by fully circularly polarized light, elliptical and bicircular fields support rescattering while at the same time in these fields the polarization vector rotates during a fraction of a cycle which can give rise to spin polarization \cite{Ayuso17}.

\acknowledgments   This work was funded by the Deutsche Forschungsgemeinschaft. We thank Olga Smirnova, Ingo Barth, Felipe Morales, and Misha Ivanov for many enlightening discussions.

\bibliographystyle{apsrev4-1}

\end{document}